\documentclass[aps,prl,showpacs, floats, twocolumn,superscriptaddress]{revtex4}
\usepackage{graphicx}  
\usepackage{dcolumn}   
\usepackage{bm}        
\usepackage{amssymb}   
\usepackage{amsmath}   
\usepackage{mathptmx}
\usepackage{xspace}
\usepackage{url}
\usepackage[colorlinks=false,linkcolor=black, citecolor=green, linktoc=page, urlcolor=cyan, 
pdffitwindow=false, draft]{hyperref}
\usepackage{footnote}
\usepackage{float}
\usepackage{multirow}
\usepackage{bm} 

\def\reff@jnl#1{{\rm#1\/}}
\def\aj{\reff@jnl{AJ}}         
\def\araa{\reff@jnl{ARA\&A}}      
\def\apj{\reff@jnl{ApJ}}        
\def\apjl{\reff@jnl{ApJ}}        
\def\apjs{\reff@jnl{ApJS}}       
\def\aap{\reff@jnl{A\&A}}        
\def\aapr{\reff@jnl{A\&A~Rev.}}     
\def\aaps{\reff@jnl{A\&AS}}       
\def\mnras{\reff@jnl{MNRAS}}      
\def\physrep{\reff@jnl{Physics Reports}}
\def\prd{\reff@jnl{Phys.Rev.D}}     
\def\prl{\reff@jnl{Phys.Rev.Lett}}   
\def\pasp{\reff@jnl{PASP}}       
\def\pasj{\reff@jnl{PASJ}}       
\def\nat{\reff@jnl{Nature}}       
\def\jcap{\reff@jnl{JCAP}}   
\def\memsai{\reff@jnl{MemSAI}} 
\def\na{\reff@jnl{New Astronomy}}       

\usepackage{color}

\def\Fref#1{Fig.~(\ref{#1})\xspace}

\def\Tref#1{Table~\ref{#1}\xspace}

\def\Eref#1{Eq.~(\ref{#1})\xspace}

\def\Cref#1{Chapter~\ref{#1}\xspace}

\def\eg{{e.g.}}
\def\ie{{i.e.}}

\def\anl{Argonne National Laboratory, 9700 South Cass Avenue, Lemont, IL 60439, USA}
\def\upenn{Department of Physics and Astronomy, University of Pennsylvania, Philadelphia, PA 19104, USA}
\def\ethz{Department of Physics, ETH Zurich, Wolfgang-Pauli-Strasse 16, CH-8093 Zurich, Switzerland}
\def\ports{Institute of Cosmology \& Gravitation, University of Portsmouth, Portsmouth, PO1 3FX, UK}
\def\ucl{Department of Physics \& Astronomy, University College London, Gower Street, London, WC1E 6BT, UK}
\def\bnl{Brookhaven National Laboratory, Bldg 510, Upton, NY 11973, USA}
\def\fermilab{Fermi National Accelerator Laboratory, P. O. Box 500, Batavia, IL 60510, USA}
\def\stanford{Department of Physics, Stanford University, 382 Via Pueblo Mall, Stanford, CA 94305, USA}
\def\kipac{Kavli Institute for Particle Astrophysics \& Cosmology, P. O. Box 2450, Stanford University, Stanford, CA 94305, USA}
\def\slac{SLAC National Accelerator Laboratory, Menlo Park, CA 94025, USA}
\def\ifae{Institut de F\'{\i}sica d'Altes Energies, Universitat Aut\`onoma de Barcelona, E-08193 Bellaterra, Barcelona, Spain}
\def\ieec{Institut de Ci\`encies de l'Espai, IEEC-CSIC, Campus UAB, Facultat de Ci\`encies, Torre C5 par-2, 08193 Bellaterra, Barcelona, Spain}
\def\ccap{Center for Cosmology and Astro-Particle Physics, The Ohio State University, Columbus, OH 43210, USA}
\def\ohio{Department of Physics, The Ohio State University, Columbus, OH 43210, USA}
\def\manchester{Jodrell Bank Center for Astrophysics, School of Physics and Astronomy, University of Manchester, Oxford Road, Manchester, M13 9PL, UK}

\def\lina{Laborat\'orio Interinstitucional de e-Astronomia - LIneA, Rua Gal. Jos\'e Cristino 77, Rio de Janeiro, RJ - 20921-400, Brazil}
\def\on{Observat\'orio Nacional, Rua Gal. Jos\'e Cristino 77, Rio de Janeiro, RJ - 20921-400, Brazil}

\def\michigan{Department of Physics, University of Michigan, Ann Arbor, MI 48109, USA}
\def\michiganastro{Department of Astronomy, University of Michigan, Ann Arbor, MI 48109, USA}
\def\chicagokavli{Kavli Institute for Cosmological Physics, University of Chicago, Chicago, IL 60637, USA}
\def\maxplanck{Max Planck Institute for Extraterrestrial Physics, Giessenbachstrasse, 85748 Garching, Germany}
\def\munich{University Observatory Munich, Scheinerstrasse 1, 81679 Munich, Germany}
\def\lmu{Department of Physics, Ludwig-Maximilians-Universitaet, Scheinerstr. 1, 81679 Muenchen, Germany}
\def\ctio{Cerro Tololo Inter-American Observatory, National Optical Astronomy Observatory, Casilla 603, La Serena, Chile}
\def\aao{Australian Astronomical Observatory, North Ryde, NSW 2113, Australia}

\def\jpl{Jet Propulsion Laboratory, California Institute of Technology, 4800 Oak Grove Dr., Pasadena, CA 91109, USA}
\def\ciemat{Centro de Investigaciones Energ\'eticas, Medioambientales y Tecnol\'ogicas (CIEMAT), Madrid, Spain}
\def\uiuc{Department of Physics, University of Illinois, 1110 W. Green St., Urbana, IL 61801, USA}
\def\ncsa{National Center for Supercomputing Applications, 1205 West Clark St., Urbana, IL 61801, USA}
\def\ua{University of Arizona, Department of Physics, 1118 E. Fourth St., Tucson, AZ 85721, USA}
\def\jpl{Jet Propulsion Laboratory, California Institute of Technology, 4800 Oak Grove Dr., Pasadena, CA 91109, USA}
\def\sussex{Department of Physics and Astronomy, Pevensey Building, University of Sussex, Brighton, BN1 9QH, UK}
\def\cluster{Excellence Cluster Universe, Boltzmannstr.\ 2, 85748 Garching, Germany}
\def\barcelona{Instituci\'o Catalana de Recerca i Estudis Avan\c{c}ats, E-08010 Barcelona, Spain}
\def\sepnet{SEPnet, South East Physics Network, (www.sepnet.ac.uk)}

\begin{document}

\widetext
\leftline{Version 1 as of \today}
\leftline{Primary authors: Chihway Chang, Vinu Vikram, Bhuvnesh Jain}


\title{Wide-Field Lensing Mass Maps from DES Science Verification Data}

\author{C.~Chang}
\email{chihway.chang@phys.ethz.ch}
\affiliation{\ethz}
\author{V.~Vikram}
\affiliation{\anl}
\affiliation{\upenn}
\author{B.~Jain}
\affiliation{\upenn}
\author{D.~Bacon}
\affiliation{\ports}  
\author{A.~Amara}
\affiliation{\ethz}

\author{M.~R.~Becker}
\affiliation{\stanford}
\affiliation{\kipac}    
\author{G.~Bernstein}
\affiliation{\upenn}  
\author{C.~Bonnett}
\affiliation{\ifae}  
\author{S.~Bridle}
\affiliation{\manchester}   
\author{D.~Brout}
\affiliation{\upenn}    
\author{M.~Busha}
\affiliation{\stanford}
\affiliation{\kipac}    
\author{J.~Frieman}
\affiliation{\chicagokavli}
\affiliation{\fermilab}
\author{E.~Gaztanaga}
\affiliation{\ieec}
\author{W.~Hartley}
\affiliation{\ethz} 
\author{M.~Jarvis}
\affiliation{\upenn} 
\author{T.~Kacprzak}
\affiliation{\ethz} 
\author{A.~Kov\'acs}
\affiliation{\ifae} 
\author{O.~Lahav}
\affiliation{\ucl} 
\author{H.~Lin}
\affiliation{\fermilab} 
\author{P.~Melchior}
\affiliation{\ccap} 
\affiliation{\ohio} 
\author{H.~Peiris}
\affiliation{\ucl} 
\author{E.~Rozo}
\affiliation{\ua} 
\author{E.~Rykoff}
\affiliation{\kipac} 
\affiliation{\slac} 
\author{C.~S\'anchez}
\affiliation{\ifae} 
\author{E.~Sheldon}
\affiliation{\bnl}   
\author{M.~A.~Troxel}
\affiliation{\manchester} 
\author{R.~Wechsler}
\affiliation{\stanford} 
\affiliation{\kipac} 
\affiliation{\slac}  
\author{J.~Zuntz}
\affiliation{\manchester} 

\author{T.~Abbott}
\affiliation{\ctio} 
\author{F.~B.~Abdalla}
\affiliation{\ucl} 
\author{S. ~Allam}
\affiliation{\fermilab} 
\author{J.~Annis}
\affiliation{\fermilab} 
\author{A.~H.~Bauer}
\affiliation{\ieec} 
\author{A.~Benoit-L{\'e}vy}
\affiliation{\ucl} 
\author{D.~Brooks}
\affiliation{\ucl} 
\author{E.~Buckley-Geer}
\affiliation{\fermilab} 
\author{D.~L.~Burke}
\affiliation{\kipac} 
\affiliation{\slac} 
\author{D.~Capozzi}
\affiliation{\ports} 
\author{A.~Carnero~Rosell}
\affiliation{\lina} 
\affiliation{\on} 
\author{M.~Carrasco~Kind}
\affiliation{\uiuc} 
\affiliation{\ncsa} 
\author{F.~J.~Castander}
\affiliation{\ieec} 
\author{M.~Crocce}
\affiliation{\ieec} 
\author{C.~B.~D'Andrea}
\affiliation{\ports} 
\author{S.~Desai}
\affiliation{\lmu} 
\author{H.~T.~Diehl}
\affiliation{\fermilab} 
\author{J.~P.~Dietrich}
\affiliation{\lmu} 
\affiliation{\cluster}  
\author{P.~Doel}
\affiliation{\ucl} 
\author{T.~F.~Eifler}
\affiliation{\upenn}
\affiliation{\jpl}  
\author{A.~E.~Evrard}
\affiliation{\michigan} 
\author{A.~Fausti Neto}
\affiliation{\lina} 
\author{B.~Flaugher}
\affiliation{\fermilab} 
\author{P.~Fosalba}
\affiliation{\ieec} 
\author{D.~Gruen}
\affiliation{\maxplanck} 
\affiliation{\munich} 
\author{R.~A.~Gruendl}
\affiliation{\uiuc} 
\affiliation{\ncsa} 
\author{G.~Gutierrez}
\affiliation{\fermilab} 
\author{K.~Honscheid}
\affiliation{\uiuc} 
\affiliation{\ncsa} 
\author{D.~James}
\affiliation{\ctio} 
\author{S.~Kent}
\affiliation{\fermilab} 
\author{K.~Kuehn}
\affiliation{\aao} 
\author{N.~Kuropatkin}
\affiliation{\fermilab} 
\author{M.~A.~G.~Maia}
\affiliation{\lina} 
\affiliation{\on} 
\author{M.~March}
\affiliation{\upenn} 
\author{P.~Martini}
\affiliation{\ccap} 
\affiliation{\ohio} 
\author{K.~W.~Merritt}
\affiliation{\fermilab} 
\author{C.~J.~Miller}
\affiliation{\michigan} 
\affiliation{\michiganastro} 
\author{R.~Miquel}
\affiliation{\ifae} 
\affiliation{\barcelona} 
\author{E.~Neilsen}
\affiliation{\fermilab} 
\author{R.~C.~Nichol}
\affiliation{\ports} 
\author{R.~Ogando}
\affiliation{\lina} 
\affiliation{\on} 
\author{A.~A.~Plazas}
\affiliation{\bnl} 
\affiliation{\jpl} 
\author{A.~K.~Romer}
\affiliation{\sussex} 
\author{A.~Roodman}
\affiliation{\kipac} 
\affiliation{\slac} 
\author{M.~Sako}
\affiliation{\upenn} 
\author{E.~Sanchez}
\affiliation{\ciemat} 
\author{I.~Sevilla}
\affiliation{\uiuc} 
\affiliation{\ciemat} 
\author{R.~C.~Smith}
\affiliation{\ctio} 
\author{M.~Soares-Santos}
\affiliation{\fermilab} 
\author{F.~Sobreira}
\affiliation{\fermilab} 
\affiliation{\lina} 
\author{E.~Suchyta}
\affiliation{\ccap} 
\affiliation{\ohio} 
\author{G.~Tarle}
\affiliation{\lmu} 
\author{J.~Thaler}
\affiliation{\uiuc} 
\author{D.~Thomas}
\affiliation{\ports} 
\affiliation{\sepnet} 
\author{D.~Tucker}
\affiliation{\fermilab} 
\author{A.~R.~Walker}
\affiliation{\ctio} 
\date{\today}

\begin{abstract}
We present a mass map reconstructed from weak gravitational lensing shear measurements over 139 deg$^2$ 
from the Dark Energy Survey (DES) Science Verification data. The mass map probes both luminous and 
dark matter, thus providing a tool for studying cosmology. We find good agreement between the mass 
map and the distribution of massive galaxy clusters identified using a red-sequence cluster finder. 
Potential candidates for super-clusters and voids are identified using these maps. 
We measure the cross-correlation between the mass map and a magnitude-limited foreground galaxy sample 
and find a detection at the 6.8$\sigma$ level with 20 arcminute smoothing. These measurements are consistent 
with simulated galaxy catalogs based on $\Lambda$CDM N-body simulations, suggesting low systematics 
uncertainties in the map. We summarize our key findings in this letter; the detailed methodology and tests for 
systematics are presented in a companion paper. 
\end{abstract}

\maketitle

\section{Introduction}

Gravitational lensing refers to the bending of light due to the curvature of space-time induced by massive 
bodies \citep{1936Sci....84..506E}. This effect allows one 
to probe the total matter distribution in the Universe, including both luminous and dark matter. Weak lensing 
is the technique of using the subtle gravitational lensing effect of a large number of galaxies to statistically infer 
the large-scale matter distribution in the Universe \citep[see][for detailed reviews]{bartelmann01, Hoekstra2008}. 
The measurement is based on small, percent-level ``shears'', or distortions of galaxy shapes due to lensing. With 
several ongoing large optical surveys collecting data \citep{2004AN....325..636H, 2005IJMPA..20.3121F, 
2012SPIE.8446E..0ZM, 2013ExA....35...25D}, this technique is one of the most powerful probes for 
constraining the nature of dark energy \citep{2006astro.ph..9591A}. 

Conventional weak lensing analyses involve calculating the N-point statistics of the shear field. In 
particular, the \textit{cosmic shear} measurement, which refers to the 2-point correlation function 
of the shear field in configuration space, has been measured 
in several earlier datasets \citep{Jarvis2006, Lin2012, Heymans2012, Jee2013, Fu2014, 
2013MNRAS.430.2200K}. 
Shear $\bm{\gamma}$ is defined to be a combination of second derivatives of the lensing potential $\psi$, 
\begin{equation}
\bm{\gamma} =\gamma_1 + i \gamma_2 = \frac{1}{2}\left(\psi_{,11} - \psi_{,22}\right) + i \psi_{,12},
\label{eq:shear}
\end{equation}
``$\psi_{,ij}=\partial^2\psi/\partial\theta_i\partial\theta_j$" is the
second partial derivative with respect to the angular sky coordinates
$\theta_i$ of $\psi$ (assuming a spatially flat Universe in the Newtonian limit of GR), which is defined as \citep{2015PhR...568....1J}
\begin{equation}
\psi\left(\bm{\theta}, r\right) = 2 \int_0^r{\mathrm{d}r^\prime \frac{r - r^\prime}{rr^{\prime}} \Phi\left(\bm{\theta}, r^\prime\right)}.
\label{eq:2dpotential}
\end{equation}
In the above equation, $r$ is the comoving distance and $\Phi$ is the 3D gravitational potential, whose spatial   
structure and time evolution contains cosmological information.

Instead of measuring statistics based on shear, here we focus on an alternative approach by converting 
shear into the projected density field, the convergence $\kappa$, also a combination of second 
derivatives of $\psi$,
\begin{equation}
\kappa = \frac{1}{2}\nabla^2 \psi = \frac{1}{2}\left(\psi_{,11} + \psi_{,22}\right).
\label{eq:kappa}
\end{equation}
The convergence directly represents the integrated mass distribution, 
which can be seen by using the cosmological Poisson equation and the Limber approximation 
to re-write \Eref{eq:kappa} as \citep{bartelmann01}
\begin{equation}
\kappa(\bm{\theta},r) = \frac{3H_0^2\Omega_m}{2} \int_0^r{\mathrm{d}r^\prime \frac{r^\prime(r-r^\prime) }{r} 
\frac{\delta\left(\bm{\theta}, r^\prime\right)} {a(r^\prime)}},
\label{eq:kappa2}
\end{equation}
where $H_{0}$ is the Hubble constant today, $\Omega_{m}$ is the total matter density today, $a$ is the 
cosmological scale factor, and $\delta = (\Delta - \bar{\Delta})/\bar{\Delta}$ is the mass overdensity 
($\Delta$ and $\bar{\Delta}$ are the 3D density and mean density respectively).  
In practice we integrate over the redshift distribution of source galaxies as shown in Eq. 15 of the 
accompanying paper \citep{2015arXiv150403002V}.

Note that the same weak lensing effect also introduces distortions in the observed cosmic microwave 
background (CMB) maps. Reconstructing the convergence map from the CMB gives the integrated 
mass up to the surface of last scattering ($z\sim1100$). Compared to the weak lensing convergence 
map constructed from galaxies, the CMB convergence map typically covers a larger area with lower 
spatial resolution, and the sources of the lensing effect (the CMB photons) come from a single 
redshift plane \citep{2014A&A...571A..17P, 
2013ApJ...771L..16H, 2011PhRvL.107b1301D}. In this letter, we use  ``weak lensing mass maps'' 
to refer to convergence maps generated from source galaxies. 

Weak lensing mass maps supplement measurements based on shear in many ways. Mass maps can be 
easily cross-correlated with other data since they represent a scalar, the local (projected) mass density, 
while the shear is a complex variable and is sensitive to the global mass distribution. Cross correlating with 
X-ray and Sunyaev-Zel'dovich observations helps us understand the relation of hot gas and dark matter in 
galaxy clusters. Cross correlating with the CMB convergence map 
provides an important cross check of lensing measurements using different tracers. Other applications 
of mass maps include peak statistics \citep{Jain2000, 2008MNRAS.391..435F, 2010MNRAS.402.1049D, 
Kratochvil2010, 2010ApJ...712..992B}, higher-order moments of $\kappa$ \citep{Waerbeke2014}, and the 
identification of superclusters and cosmic voids \citep{heymans08}.   

The methodology of generating weak lensing mass maps has been demonstrated in earlier work. 
\citet{2007Natur.445..286M} generated a 3D mass map using COSMOS data in a 1.64 deg$^{2}$ 
area. The high-quality shear measurements and redshift information allow for good mass 
reconstruction on small scales and in the radial direction. \citet{waerbeke13}, on the other hand, focused 
on larger-scale information and generated 2D wide-field mass maps from four fields of size 25--72 deg$^{2}$ 
in the Canada-France-Hawaii Telescope Lensing Survey (CFHTLenS). Our work is similar to 
\citet{waerbeke13}, but uses one contiguous region of 139 deg$^{2}$ from the Dark Energy 
Survey \citep[DES,][]{DES2005, 2005IJMPA..20.3121F} data. This is the first step towards building mass 
maps from the full DES data set. 

The data used in this work is part of the Science Verification (SV) dataset from DES,  
an ongoing  ground-based galaxy survey that is scheduled to operate from September 2013 to 
February 2018. The SV data were collected between November 2012 and February 2013 shortly after the 
commissioning of the new wide-field mosaic camera, the Dark Energy Camera 
\citep[DECam,][]{2012PhPro..37.1332D, 2012SPIE.8446E..11F, 2015arXiv150402900F} on the 4m Blanco 
telescope at the Cerro Tololo Inter-American Observatory (CTIO) in Chile. 
This data was used to test survey operations and assess data quality. The images are taken in 5 optical filter 
bands ($grizY$) on a total area of $\sim250$ deg$^{2}$ and reach close to the expected full depth of 
DES at $r\sim 23.9$.
 
The main goal of this work is to reconstruct the weak lensing mass map from shear measurements of 
the DES SV data in a 139 deg$^{2}$ contiguous region overlapping with the South Pole Telescope 
Survey (the SPT-E field). We present the methodology used for the map construction, followed by 
cross-correlation results and conclusions. Throughout the paper, we adopt the following cosmological 
parameters: 
$\Omega_{m}=0.3$, 
$\Omega_{\Lambda}=0.7$, $\Omega_{k}=0.0$, $h=0.72$. 
A detailed account of this work can be found in a companion paper in PRD \citep{2015arXiv150403002V}.  

\section{Methodology}
\subsection{Data and simulations}

Our galaxy samples are based on the DES SV Gold catalog (Rykoff et al., in preparation) and several 
extensions to it. The Gold catalog is a product of the DES Data Management \citep[DESDM,][]{2006SPIE.6270E..23N, 
2011arXiv1109.6741S, 2012ApJ...757...83D, 2012SPIE.8451E..0DM} pipeline version 
``SVA1'' (Yanny et al., in preparation), which includes calibrated photometry and astrometry, object morphology, 
object classification and masking of the co-add SV images. DESDM utilizes the software packages 
\textsc{SCAMP} \citep{2006ASPC..351..112B}, \textsc{SWARP} \citep{2002ASPC..281..228B}, 
\textsc{PSFEx} \citep{2011ASPC..442..435B} and \textsc{SExtractor} \citep{1996A&AS..117..393B} in the pipeline. 

Several additional catalogs are used in this work. 
We use a photometric redshift (photo-$z$) catalog from the photo-$z$ code 
Bayesian Photometric Redshifts \citep[BPZ,][]{2000ApJ...536..571B,2006AJ....132..926C}. 
We use two shear catalogs from the \texttt{ngmix} 
code \citep{2014MNRAS.444L..25S} and the 
\texttt{im3shape} 
code \citep{2013MNRAS.434.1604Z}. 
The two independent shear catalogs allows us to assess the robustness of the measurement. 
The shear measurement algorithms operate on single-exposure images and measure the galaxy shapes, 
or ``ellipticities'', by jointly fitting the images of the same galaxies obtained in different exposures
with one galaxy model and the 
different point-spread-function (PSF) model in each image. The resulting ellipticity is a noisy estimator for 
shear \citep{bartelmann01}. The shear estimates used in this work have been tested rigorously as described 
in Jarvis et al. (in preparation). 

\begin{table}
\begin{center}
\caption{Catalogs and selection criteria used to construct the foreground and background sample for this 
work, and the number of galaxies in each sample after all the cuts are applied. 
The redshift cut is based on the mean redshift output from the BPZ photo-$z$ code and the magnitude 
cut is based on the \texttt{MAG\_AUTO} parameter in the \textsc{SExtractor} output.}
\begin{tabular}{lcccc}
\\
\hline 
& \multicolumn{2}{c}{Background (source)} & Foreground (lens)\\ \hline
Input catalog & \texttt{ngmix011}   & \texttt{im3shape}    & SVA1 Gold    \\ \hline   
Photo-$z$ & \multicolumn{2}{c}{0.6$<$z$<$1.2} & 0.1$<$z$<$0.5  \\
Selection &  \multicolumn{2}{c}{``conservative additive'' }& $i<$22 \\ 
Number of galaxies &  1,111,487 & 1,013,317 &  1,106,189   \\   
Number density & \multirow{2}{*}{2.22}  & \multirow{2}{*}{2.03} &  \multirow{2}{*}{2.21}   \\  
(arcmin$^{-2}$) &   &   &   \\
Mean redshift   &  0.826 & 0.825  & 0.367  \\  
\hline
  \vspace{-0.5in}
\end{tabular}
\label{tab:sample_selection}
\end{center}
\end{table}

We extract from these catalogs background (``source'') and foreground (``lens'') galaxy samples. The objective 
is to construct the convergence, or mass map, from the background sample and cross-correlate it with the 
weighted galaxy map built from the foreground sample. \Tref{tab:sample_selection} lists the final selection 
criteria for the samples. The foreground sample is magnitude-limited at $i=22$, while the background sample 
is selected through a series of lensing tests (Jarvis et al. in preparation) and is not complete. The incompleteness 
of the background sample affects only the spatial distribution of the noise on these maps but does not bias the 
signal. In the companion paper we describe in detail the construction of these samples and 
also discuss a second foreground sample composed of luminous red 
galaxies (LRGs). 
Note that the plots in this letter rely on the \texttt{ngmix} shear catalog. However 
we analyzed both shear catalogs to assess their
statistical consistency. The ``conservative additive'' selection criteria on the background sample involves a 
combination of signal-to-noise (S/N) cuts, size cuts and other quality cuts. 

To facilitate our understanding of possible systematics in the procedure of constructing the mass map, we use 
a set of simulated galaxy catalogs that we match closely to the characteristics of the data (including intrinsic 
galaxy properties, galaxy number counts, noise, photo-$z$ errors, survey mask). We use the simulated galaxy 
catalogs 
developed for the DES collaboration \citep{2013AAS...22134107B}. The catalog is 
based on three flat $\Lambda$CDM dark matter-only N-body simulations with different resolutions. Galaxies 
are populated using the prescriptions derived from a high-resolution simulation using SubHalo Abundance 
Matching techniques \citep{2006ApJ...647..201C, 2013ApJ...771...30R, 2013AAS...22134107B}. 
Photometric properties for each galaxy are then assigned so that the magnitude-color-redshift distribution 
reproduces that observed in the SDSS DR8 \citep{2011ApJS..193...29A} and DEEP2 \citep{2013ApJS..208....5N} 
data. Weak lensing parameters (shear and 
convergence) are assigned to each galaxy based on the high-resolution ray-tracing algorithm Curved-sky 
grAvitational Lensing for Cosmological Light conE simulatioNS \citep{2013MNRAS.435..115B}. 
Details of the data and simulation catalogs are presented in the companion paper.

\begin{figure*}
  \begin{center}
  \includegraphics[scale=0.53]{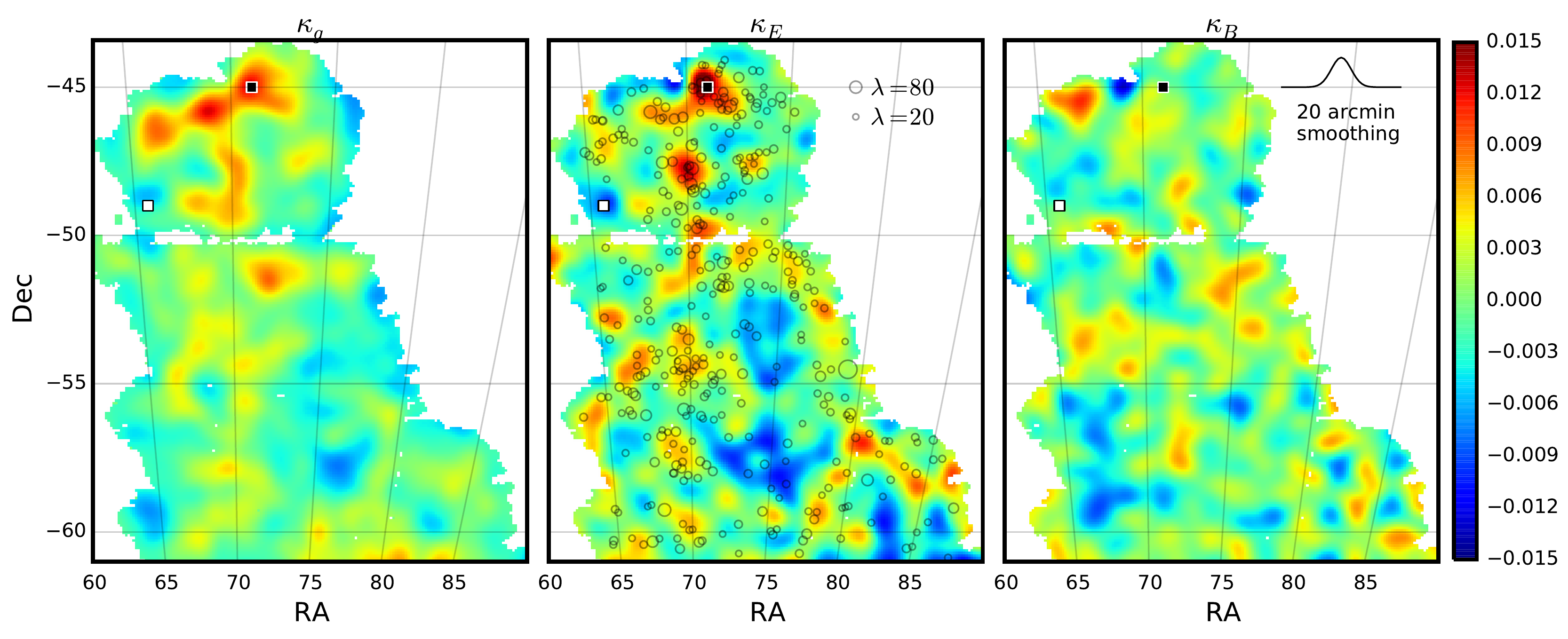}  
  \vspace{-0.4in}
  \end{center}
  \caption{The DES SV weighted foreground galaxy maps $\kappa_{g, main}$ (left), E-mode convergence map 
  $\kappa_{E}$ (middle) and B-mode convergence map $\kappa_{B}$ (right) are shown in these panels. 
  All maps are generated with 5$\times$5 arcmin$^{2}$ pixels and 20 arcmin RMS Gaussian smoothing. In 
  the $\kappa_{g}$ and $\kappa_{E}$ maps, red areas corresponds to overdensities and blue areas to 
  underdensities. White regions correspond to the survey mask. The scale of the Gaussian smoothing kernel 
  is indicated by the Gaussian profile on the upper right corner of the right panel. 
  The $\kappa_{E}$ map is overlaid by Redmapper galaxy clusters with optical richness $\lambda>20$. The 
  radius of the circles scale with $\lambda$. The black and white squares show the super cluster and super void 
  candidate we investigate in \Fref{fig:supercluster}.}
\label{fig:massmap}
\end{figure*}

\subsection{Mass and weighted galaxy maps}
\Eref{eq:shear} and \Eref{eq:kappa} 
can be Fourier transformed to get a simple relationship between the Fourier transforms of the 
shear and convergence, denoted $\hat{\bm{\gamma}}$ and $\hat{\kappa}$ \citep{ks93}:
\begin{equation}
\hat{\kappa}_{\bm{\ell}}= D^*_{\bm{\ell}} \hat{\bm{\gamma}}_(\bm{\ell}),
\label{eq:ks_ft}
\end{equation}
\begin{equation}
D_{\bm{\ell}} = \frac{\ell_1^2 - \ell_2^2 + 2 i \ell_1 \ell_2}{|\bm{\ell}|^2},
\label{eq:D}
\end{equation}
where $\ell_i$ are the components of the angular wavenumber. 
The above equations hold  for $\bm{\ell}> 0$. 

In practice, we pixelate the shear measurements into a map of 5$\times$5 arcmin$^{2}$ pixels and 
Fourier transform the map. We then use \Eref{eq:ks_ft} to obtain $\hat{\kappa}$ and inverse Fourier 
transform to yield our final real-space convergence map. In an ideal scenario, this reconstructed 
convergence map does not contain an imaginary component. However, due to noise, the finite area 
of the map, and masking, a non-zero imaginary component is recovered. We separate the real and 
imaginary parts of the measured convergence map into E- and B-modes, or 
$\kappa = \kappa_{E} + i \kappa_{B}$. The B-mode convergence is a useful diagnostic tool for testing 
systematics, as it should vanish for real lensing signals on a sufficiently large area. Finally, as the 
uncertainty in this reconstruction is formally infinite for a discrete set of noisy shear estimates, it is 
important to apply a filter to remove the high-frequency noise \citep{waerbeke00}. In this work we apply 
a Gaussian filter of different sizes. In the companion paper \citep{2015arXiv150403002V} 
we use simulations to quantify the degradation in $\kappa_{E}$ and the level of $\kappa_{B}$ 
expected from the noise and masking in the data. We find that our results are consistent with that 
expected from simulations.

One of the main goals of this work is to cross-correlate the mass map with the foreground galaxy 
distribution. For this purpose, we construct a second mass map assuming that the foreground galaxy 
sample traces the mass distribution --  we refer to this map as $\kappa_{g}$. It is constructed using  
equation \Eref{eq:kappa2} with $\delta$ replaced by $\delta_{g}$, the fractional overdensity of galaxy 
counts. Under the assumption of linear bias (\ie\ galaxy overdensities are linearly proportional to the 
total mass overdensities, which is expected to be valid on sufficiently large scales), the smoothed
$\kappa_{g}$ is simply a product of the mass map $\kappa$ with a constant bias factor. For 
our foreground galaxy sample, the linear bias is valid above 5-10 arcmin scales, which is the 
focus of our study \citep{2007A&A...461..861S}.
In practice, the limited redshift range of our foreground galaxy sample means that we cannot expect 
a perfect estimate of the mass map even if the bias factor were unity. 

\section{Results}
\Fref{fig:massmap} shows the resulting weighted galaxy map and the E and B-mode convergence maps 
generated from the procedure described above. The maps shown are for a Gaussian smoothing of 
20 arcmin RMS. We expect $\kappa_{E}$ to correlate with $\kappa_{g}$, while $\kappa_{B}$ should 
not correlate with either of the other maps. 

\subsection{Correlation with clusters}
The $\kappa_{E}$ map shown in the middle panel of \Fref{fig:massmap} is overlaid with galaxy 
clusters detected in the same data using the algorithm Redmapper 
\citep{2012ApJ...746..178R}.
Each cluster is represented by a circle with radius proportional to the 
optical richness $\lambda$, which is related to mass via a roughly linear relation (see 
\citet{2012ApJ...746..178R} for details 
of the mass calibration of $\lambda$). We select only clusters with $\lambda>20$, which corresponds 
to mass $\gtrsim 1.7 \times 10^{14}$ M$_{\odot}$ ($\lambda=80$ corresponds to mass 
$\sim 7.6 \times 10^{14}$ M$_{\odot}$). Visually, one can see that the spatial 
distribution of the clusters traces the mass map very well, with most clusters detected in or around the high
$\kappa_{E}$ regions. 

\begin{figure}
  \begin{center}
  \includegraphics[scale=0.46]{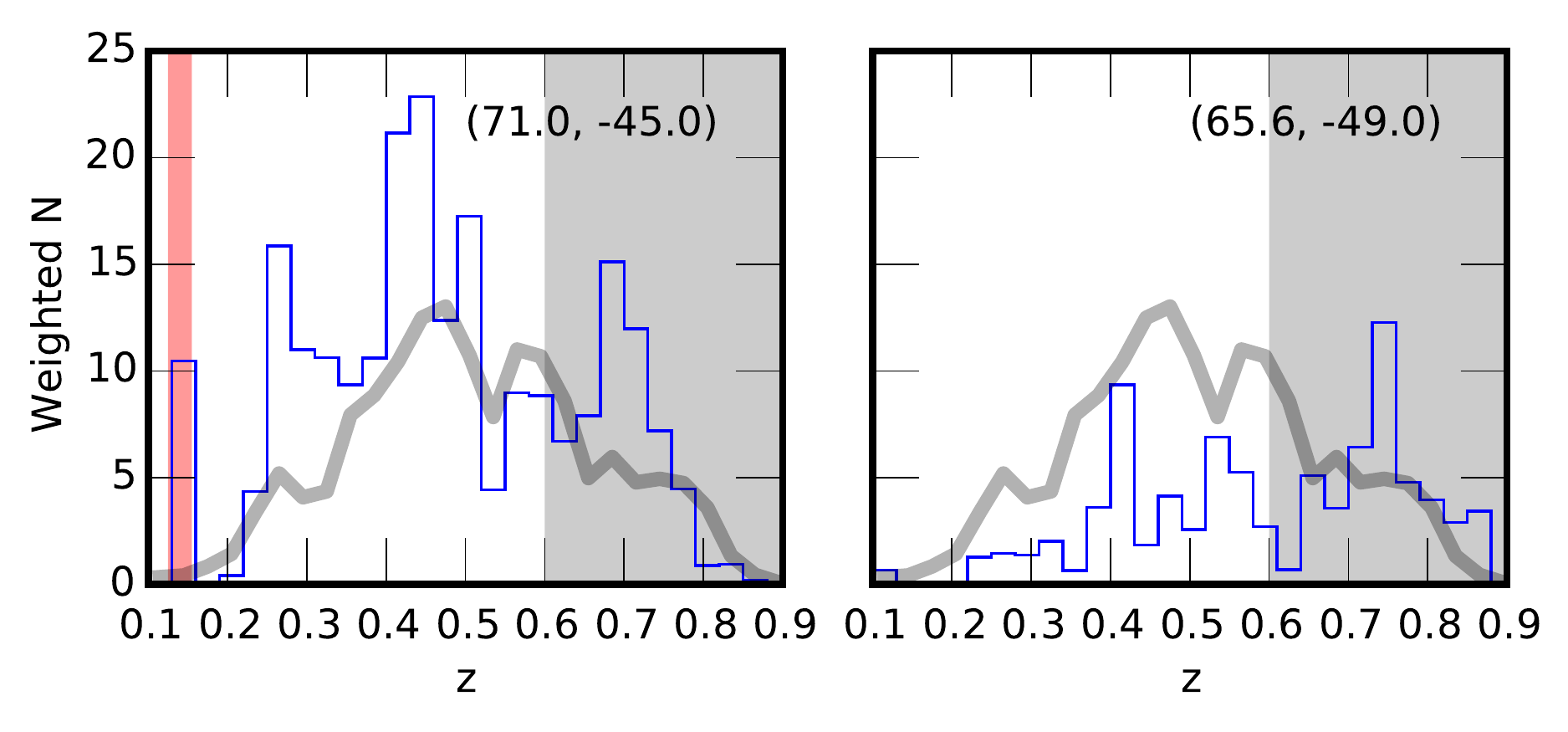}  
    \vspace{-0.4in}
  \end{center}
  \caption{Blue lines show the richness-weighted redshift distribution of redmapper galaxy clusters along  
  along overdense (left) and underdense (right) regions in the convergence map marked by the black and white 
  squares in \Fref{fig:massmap}. The 
  (RA, Dec) positions of the each region is shown in the upper right corner of each panel. The thick grey 
  line shows the average redshift distribution over the full map. Both lines are weighted by the lensing efficiency. 
  The redshift range above z = 0.6 (marked with the shaded grey area) overlap with the background sample, hence 
  the interpretation of the structures there is more complicated.}
\label{fig:supercluster}
\end{figure}

We analyze the redshift distributions of the clusters in the high and low mass density regions. Two examples 
are shown in \Fref{fig:supercluster}, where we plot (in blue) the lensing efficiency and $\lambda$-weighted redshift 
distribution of the clusters within a 1 degree radius of the identified high and low-mass positions marked in 
\Fref{fig:massmap}. Compared to the average redshift distribution of clusters (overlaid in grey), we find that the 
high-mass (low-mass) regions indeed contain many more (fewer) clusters than average. The redshift binning is 
$\Delta z$=0.03, corresponding to between 1.5--3 $\sigma_{z}$ in this redshift range, where $\sigma_{z}$ 
is the cluster photo-$z$ error uncertainty. The photo-$z$'s for Redmapper clusters are very well 
determined ($\sigma_z \approx 0.01(1+z)$), which is important for the identifications of the 3D structures. 
Using these histograms we can identify potential candidates for super-clusters. For example, the peak at 
$z\sim0.14$ in the left panel parked in red indicates that this spatial structure is contained in a redshift range 
localized to within about 100 Mpc along the line of sight. This line of sight has multiple structures at different 
redshifts, others have just one or two. The redshift range above $z=0.6$ (marked with the shaded grey area) 
overlaps with the background sample, hence the interpretation of their relation with the 
mass map is more complicated. The largest mass concentrations are investigated in more detail in the 
companion paper and in follow-up studies. 

\subsection{Mass-galaxy correlation}

Next, we investigate quantitatively the correlation between the foreground galaxies and the mass 
map by calculating the Pearson correlation coefficient between the two maps over a range of 
smoothing scales that span 5 to 40 arcmin. That is, we calculate  
\begin{equation}
\rho_{\kappa_{E} \kappa_g} = \frac{\langle\kappa_{E}  \kappa_g\rangle}{\sigma_{\kappa_{E}} \sigma_{\kappa_g}},
\label{eq:cc}
\end{equation}
where $\langle\kappa_{E} \kappa_g\rangle$ is the covariance between $\kappa_{E}$ and $\kappa_g$, and 
$\sigma_{\kappa_{E}}$ and $\sigma_{\kappa_g}$ are the standard deviations of the two maps. 
In this calculation, pixels in the masked region are not used. We also remove pixels within 10 arcmin of the 
boundaries to avoid significant artefacts from the smoothing. Similarly we calculate the Pearson correlation 
coefficient between $\kappa_{B}$ and the other maps to check for any significant systematic effects. The errors 
on the correlation coefficients are estimated by a jackknife resampling of 10 deg$^{2}$ sub-regions of the maps 
(each jackknife subsample is $\sim93\%$ of the total area). 

\begin{figure}
  \begin{center}
  \includegraphics[scale=0.37]{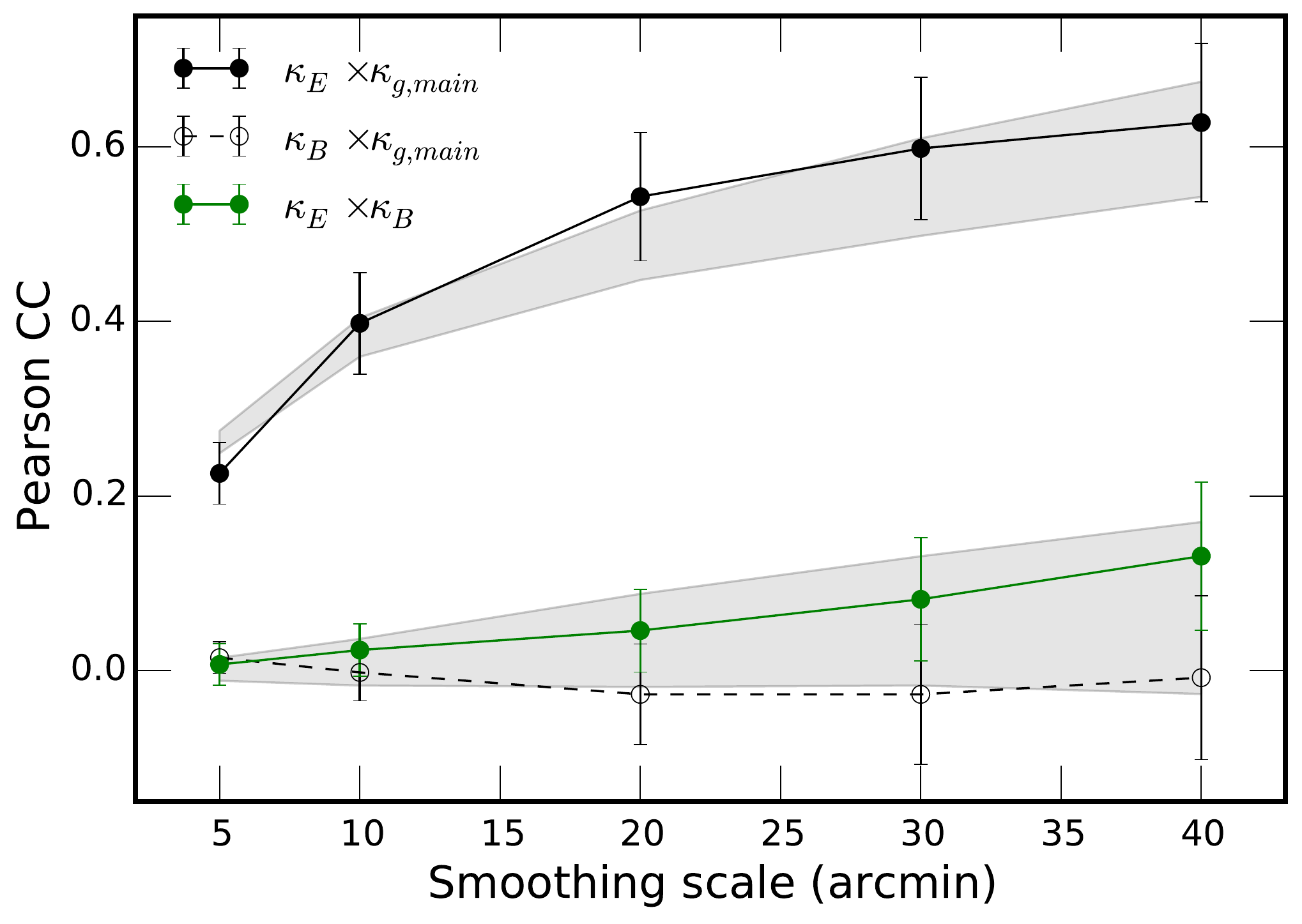}  
   \vspace{-0.2in}
  \end{center}
  \caption{The Pearson correlation coefficient between the foreground galaxy and 
   convergence maps is shown as a function of smoothing scale. The solid and open symbols show correlation 
  coefficients from the E  and B-modes of the convergence respectively. 
  The grey shaded regions show the 
  $1\sigma$ bounds from simulations for the correlation between the E and B-mode convergence and the 
  foreground galaxies, with the same pixelization and smoothing as the data as well as sources of statistical uncertainty. 
  The green points show the correlation between E and 
  B-modes of the convergence map. The various correlation coefficients with the B-mode convergence 
  are consistent with zero. Uncertainties on all measurements are estimated using jackknife resampling. }
\label{fig:cc_vs_scale}
\end{figure}

The results are shown in \Fref{fig:cc_vs_scale}. We find that the Pearson correlation coefficient between 
$\kappa_{g}$ 
and the E-mode convergence 
is $0.39\pm 0.06$ 
at 10 arcmin smoothing and 
$0.52 \pm 0.08$ 
at 20 arcmin smoothing.
This corresponds to a $\sim 6.8 \sigma$ 
significance 
at these scales. The correlation between the B-mode 
convergence and the $\kappa_{g}$ maps is consistent with zero at all smoothing scales. The 
correlation between the E and B-modes convergence is also consistent with zero. 
The grey shaded regions show the 1$\sigma$ range of results from the simulated galaxy catalogs 
modelled to match the main characteristics of the data samples. The black data points agree well 
with the simulations, suggesting there are no significant contributions to our signal from systematic errors.

To further examine the potential contamination by systematics in the maps, we construct maps of 20 
quantities associated with the observing conditions (\eg\ airmass, extinction, seeing, PSF ellipticity 
etc.) and cross correlate with our 
$\kappa_{E}$ and $\kappa_{g}$ maps. We find that none of these quantities contribute 
significantly to the cross correlation signal we have measured, with most of them consistent 
with zero. Details are presented in the companion paper. 

\section{Summary}
We present in this letter a weak lensing convergence map generated from shear measurements in the 
139 deg$^{2}$ SPT-E field in the Dark Energy Survey Science Verification data. The mean redshift 
of the source galaxies is $0.82$ and corresponds to a comoving distance of 2.9 Gpc. This map probes the 
projected total mass (luminous and dark), with matter approximately half-way between us and the
source galaxies 
making 
the most contribution to the lensing. We study the correlation of the mass map with galaxies and clusters that 
trace the foreground mass distribution. 

The spatial distribution of galaxy clusters identified in the same data using an independent technique is highly 
correlated with the mass map. The combination of the mass map and the cluster catalog provide a powerful tool for 
exploring potential super-clusters and super-voids in the Universe. Cross-correlating the E-mode mass map with 
a magnitude-limited foreground galaxy sample gives a 6.8$\sigma$ detection at 20 arcminute smoothing, while the 
cross correlation between B-mode mass map and the galaxies is consistent with zero on all scales. The 
cross-correlation between E and B-mode mass map are also consistent with zero. These results are consistent 
with simulations of the $\Lambda$CDM model in which we have modeled several sources of statistical 
uncertainties in the lensing and weighted galaxy maps. More detailed analysis, simulation and systematics tests 
are described in a companion PRD paper \citep{2015arXiv150403002V}.

Topics for 
follow-up studies include the study of galaxy bias, identification of super-clusters and super-voids, higher 
order moments of the mass map, and cross-correlation with the CMB and other observations. With the full 
set of data from DES in a few years ($\sim 35$ times the size of the SV data used in this work), we expect 
the mass maps to be a powerful tool for cosmology. 

\section*{Acknowledgements}

We are grateful for the extraordinary contributions of our CTIO colleagues and the DECam 
Construction, Commissioning and Science Verification teams in achieving the excellent 
instrument and telescope conditions that have made this work possible. The success of this 
project also relies critically on the expertise and dedication of the DES Data Management group.

We thank Jake VanderPlas, Andy Connolly, Phil Marshall, Ludo van Waerbeke and Rafal Szepietowski 
for discussions and collaborative work on mass mapping methodology. 
CC and AA are supported by the Swiss National Science Foundation grants 
200021-149442 and 200021-143906. SB and JZ acknowledge support from a 
European Research Council Starting Grant with number 240672. 
DG was supported by
SFB-Transregio 33 `The Dark Universe' by the Deutsche
Forschungsgemeinschaft (DFG) and the DFG cluster of excellence `Origin
and Structure of the Universe'. 
FS acknowledges financial support provided by CAPES under contract No. 3171-13-2. 
OL acknowledges support from a European Research Council Advanced Grant FP7/291329

Funding for the DES Projects has been provided by the U.S. Department of Energy, the U.S. National Science 
Foundation, the Ministry of Science and Education of Spain, the Science and Technology Facilities Council of 
the United Kingdom, the Higher Education Funding Council for England, the National Center for Supercomputing 
Applications at the University of Illinois at Urbana-Champaign, the Kavli Institute of Cosmological Physics 
at the University of Chicago, the Center for Cosmology and Astro-Particle Physics at the Ohio State University,
the Mitchell Institute for Fundamental Physics and Astronomy at Texas A\&M University, Financiadora de 
Estudos e Projetos, Funda{\c c}{\~a}o Carlos Chagas Filho de Amparo {\`a} Pesquisa do Estado do Rio de 
Janeiro, Conselho Nacional de Desenvolvimento Cient{\'i}fico e Tecnol{\'o}gico and the Minist{\'e}rio da 
Ci{\^e}ncia e Tecnologia, the Deutsche Forschungsgemeinschaft and the Collaborating Institutions in the 
Dark Energy Survey. 

The DES data management system is supported by the National Science Foundation under Grant Number 
AST-1138766. The DES participants from Spanish institutions are partially supported by MINECO under 
grants AYA2012-39559, ESP2013-48274, FPA2013-47986, and Centro de Excelencia Severo Ochoa 
SEV-2012-0234, some of which include ERDF funds from the European Union.

The Collaborating Institutions are Argonne National Laboratory, the University of California at Santa Cruz, 
the University of Cambridge, Centro de Investigaciones Energeticas, Medioambientales y Tecnologicas-Madrid, 
the University of Chicago, University College London, the DES-Brazil Consortium, the Eidgen{\"o}ssische 
Technische Hochschule (ETH) Z{\"u}rich, Fermi National Accelerator Laboratory,
the University of Edinburgh, 
the University of Illinois at Urbana-Champaign, the Institut de Ci\`encies de l'Espai (IEEC/CSIC), 
the Institut de F\'{\i}sica d'Altes Energies, Lawrence Berkeley National Laboratory, the Ludwig-Maximilians 
Universit{\"a}t and the associated Excellence Cluster Universe, the University of Michigan, the National Optical 
Astronomy Observatory, the University of Nottingham, The Ohio State University, the University of Pennsylvania, 
the University of Portsmouth, SLAC National Accelerator Laboratory, Stanford University, the University of 
Sussex, and Texas A\&M University.

This paper has gone through internal review by the DES collaboration.


\end{document}
%